\documentclass[11pt,twoside]{article}
\usepackage{FUSE2004}
\usepackage{natbib}
\usepackage{epsf}
\usepackage{psfig}
\usepackage{lscape}
\begin{document}
\title{FUSE Observations of Variability in Hot-Star Winds}

\author{A.~W. Fullerton\altaffilmark{1}}
\affil{Dept. of Physics \& Astronomy,
       University of Victoria,
       P.O. Box 3055,
       Victoria, BC, V8W 3P6,
       Canada}
\altaffiltext{1}{Postal Address: Dept. of Physics \& Astronomy,
	         The Johns Hopkins University,
		 3400 N. Charles Street,
		 Baltimore, MD 21218,
		 U.S.A.}
\author{D.~L. Massa}
\affil{SGT Inc.,
       Code 681.0,
       NASA's Goddard Space Flight Center,
       Greenbelt, MD 20771,
       U.S.A.}
\author{R.~K. Prinja, I.~D. Howarth, A.~J. Willis}
\affil{Dept. of Physics \& Astronomy,
       University College London,
       Gower Street,
       London, WC1E 6BT,
       England}
       
\author{S.~P. Owocki} 
\affil{Bartol Research Institute,
        University of Delaware,
	Newark, DE 19716,
	U.S.A.}

\begin{abstract}
New constraints on the large-scale wind structures responsible for
discrete absorption components are obtained from far-ultraviolet
time-series observations of O supergiants in the Large Magellanic Cloud.
\end{abstract}

\section{What Causes DACs?}
Discrete absorption components (DACs) are optical depth enhancements 
in the absorption troughs of P~Cygni profiles, which (a)
occur very commonly in wind profiles of OB-type stars;  
(b) accelerate slowly compared with the expected wind flow; and
(c) often recur on time scales related to stellar rotation.  
Since the radial optical depth at any position $v \equiv v(r)$ in an
expanding wind is given by the Sobolev approximation 
\begin{displaymath}
   \tau_{\rm rad}(v)\propto ~ q_i ~\rho~ \left( {{\rm d}v \over {\rm d}r } \right)^{-1}, 
\end{displaymath}
such enhancements can in general be produced by   
{\em increasing} the local ion fraction, $q_i$\,;  
{\em increasing} the local wind density, $\rho$\,; 
{\em flattening} the local velocity gradient, ${\rm d}v / {\rm d}r$\,; or by 
some combination of these factors, which are inter-related.
The far-ultraviolet region of the spectrum provides access to many
different resonance lines, and consequently time-series observations
with {\it FUSE\/} provide the diagnostic capability required to disentangle 
these factors.

\section{{\it FUSE\/} Time Series Observations}

\begin{figure}
   \vspace*{-0.15in}
   {{\plotone{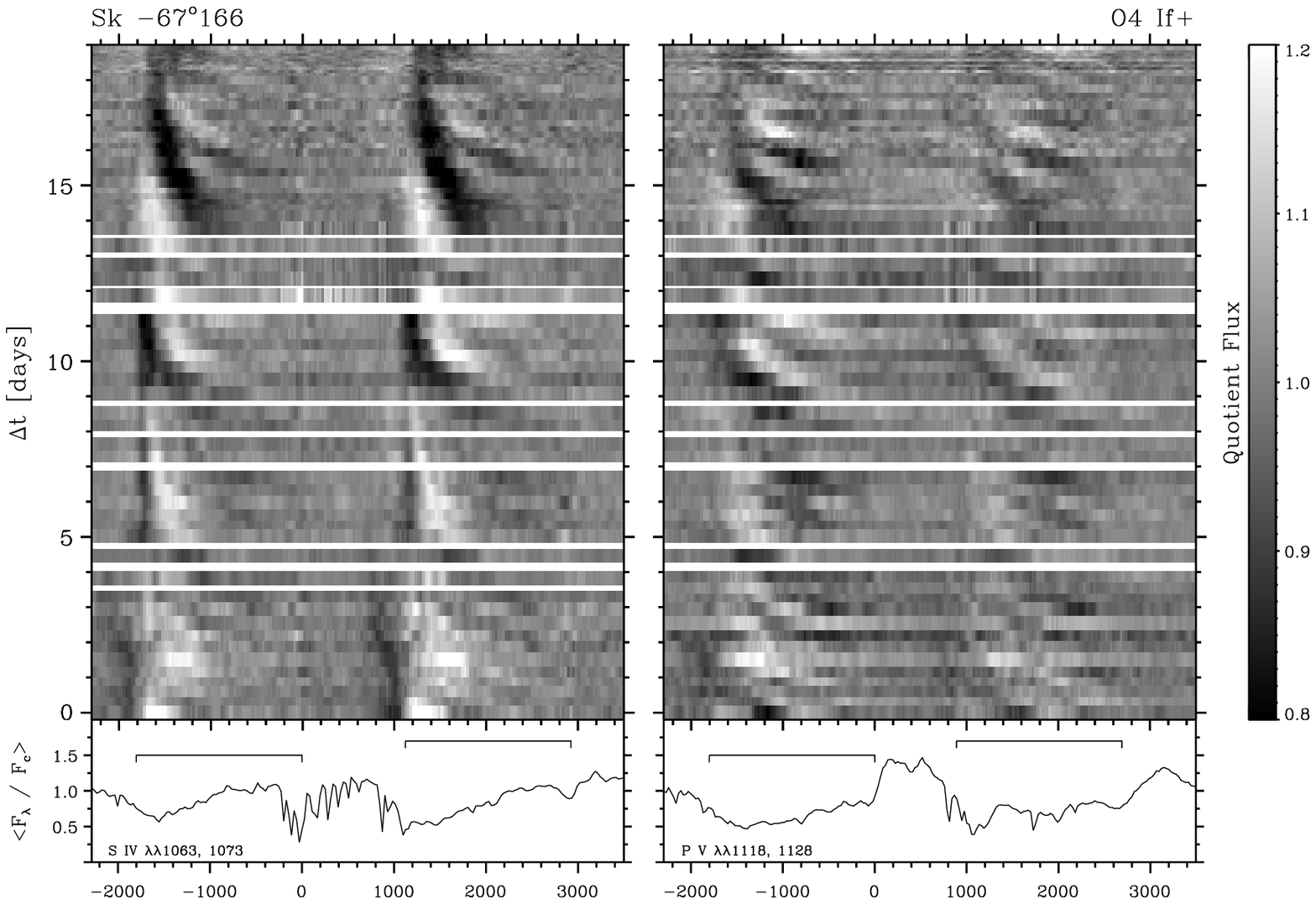}} 
   {\plotone{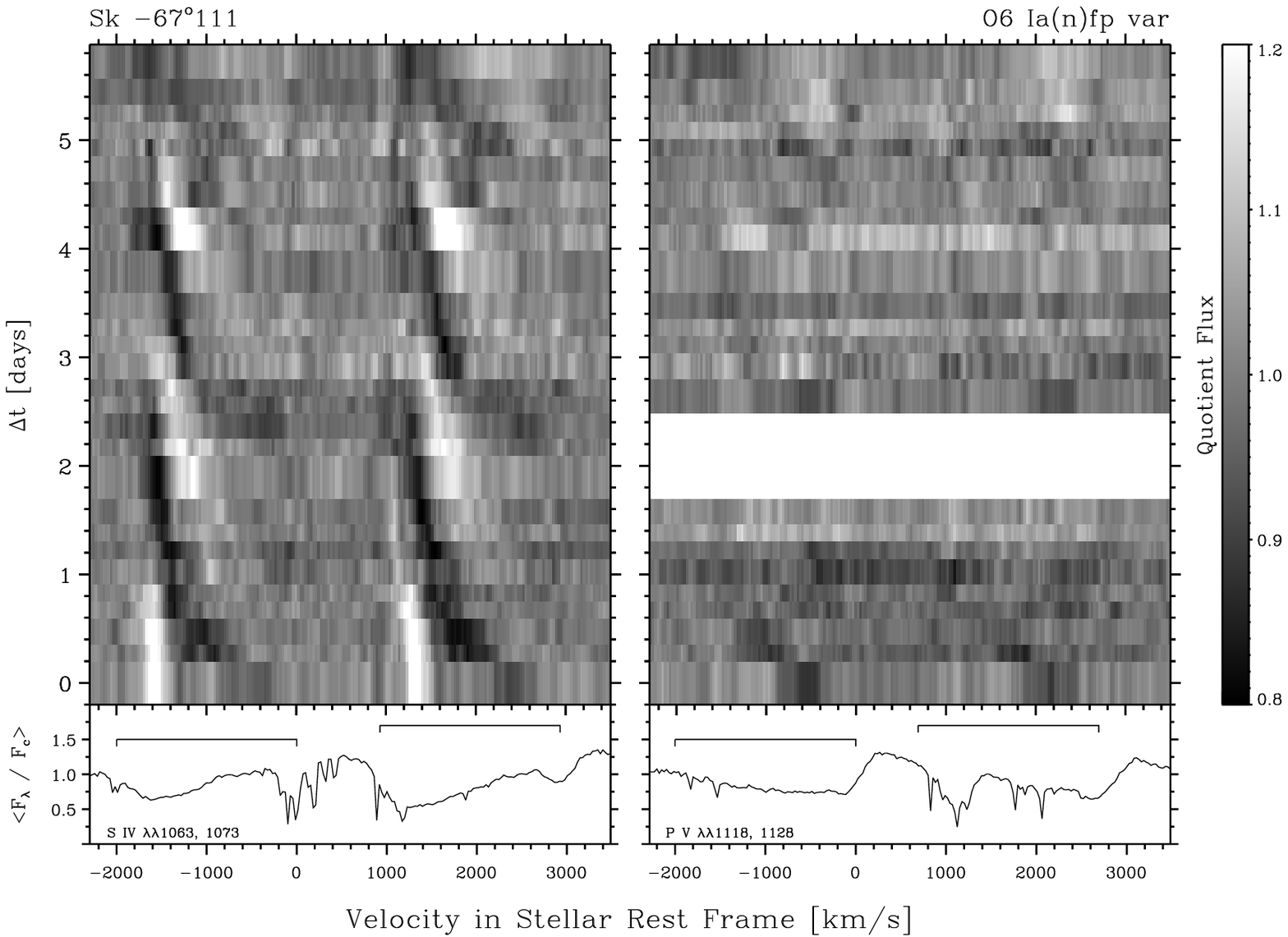}}}
   \caption{Dynamic spectra for the {\ion{S}{4}} (left) and {\ion{P}{5}} 
            (right) lines of {Sk\,$-$67\deg166} (top) and {Sk\,$-$67\deg111} 
	    (bottom).
            The dynamic spectra were formed by dividing individual
            spectra in the time series by the mean profile.
	    They are plotted with the same velocity and intensity scales to 
	    permit direct comparison, though the   
	    time axes are different for the two stars.
	    \label{dynspec}
	    }	    
\end{figure}

We have used {\it FUSE\/} to monitor two O-type supergiants in the Large
Magellanic Cloud:
{Sk\,$-$67\deg166} [O4~If+] for 18.9 days, where $P_{\rm rot} \la 13$ days; 
and
{Sk\,$-$67\deg111} [O6~Ia(n)fp~var] for 5.8 days, where 
$P_{\rm rot} \la 4.5$ days.
These time series are illustrated in Figure~\ref{dynspec}.
For both stars, DACs recur cyclically in all unsaturated resonance lines 
on time scales consistent with their estimated rotational periods.
They propagate {\em in phase} through the resonance lines of dominant ions
(e.g., {\ion{P}{5}}) and their adjacent stages
(e.g., {\ion{P}{4}}, {\ion{S}{4}}, {\ion{S}{6}}), as well
as trace species like {\ion{O}{6}}.
However, the DACs have systematically larger amplitudes in the wind profiles 
of the lower ions, and are substantially weaker in lines of dominant ions.
The DACs are barely detectable in the {\ion{P}{5}} lines of {Sk\,$-$67\deg111}.

The behavior of DACs in different ionic species permit two inferences:
\begin{enumerate}

\item Since the {\em fractional} amplitudes of DACs are different in resonance 
      lines of different ions, DACS {\em cannot} be solely due to flattening 
      of the velocity gradient or localized density enhancements.  
      The absence of variations in {\ion{N}{4}}~$\lambda$955, an excited 
      transition of the dominant ion, further indicates that any density 
      fluctuations associated with DACs are small at low velocities.

\item Since the DACs propagate in phase in the resonance lines of all 
      ionization stages where they are detected, from ions below
      the dominant stage to super-ions 
      {\citep[in particular: {\ion{S}{4}} and {\ion{S}{6}}, which exhibit
      anti-correlated abundances in smooth, steady models; 
      see, e.g.,][]{Crowther02}},
      {\em either} DACs are unlikely to be produced solely by ionization 
      changes {\em or} their ionization conditions are very different from 
      the background flow.

\end{enumerate}

These results suggest that DACs are not caused by changes in a single physical 
variable in the expression for $\tau_{\rm rad}(v)$, but must instead be 
caused by interactions between them.
Suppose, e.g., that there is a local flattening of the velocity gradient, 
as indicated by hydrodynamic simulations of ``Co-Rotating Interaction Regions" 
{\citep{Cranmer96}}.
This flattening is accompanied by an increase in $\rho$ at similar velocities, 
though different spatial locations.
The observations require that these density perturbations, together with 
changes in ${\rm d}v/{\rm d} r$, are small enough to produce only minor
variations in the wind profiles of excited transitions and dominant ions.
However, since the abundances of lower ions are preferentially 
increased by recombination, DACs would be fractionally stronger in them.
Finally, enhanced production of lower ions like {\ion{S}{4}} and 
{\ion{O}{4}} might lead to {\em correlated} production of higher ions like 
{\ion{S}{6}} and {\ion{O}{6}} via Auger ionization, particularly if sufficient 
X-ray flux is produced locally by the disturbances in the flow.
Ongoing analysis is aimed at disentangling the complicated relationships 
between these factors more rigorously.

\acknowledgements{The NASA--CNES--CSA Far Ultraviolet Spectroscopic Explorer 
is operated by the Johns Hopkins University for NASA under contract NAG-32985.}

\end{document}